\documentclass[twocolumn,showpacs,preprintnumbers,amsmath,amssymb]{revtex4}
\usepackage{graphicx}% Include figure files
\usepackage{dcolumn}% Align table columns on decimal point
\usepackage{bm}% bold math
\usepackage{colordvi}

\begin{document}

\title{Dimensional collapse and fractal attractors of a system with fluctuating delay times}% Force line breaks with \\

\author{Jian Wang}
\author{G\"unter Radons}%
 \thanks{corresponding authors}
% \email{radons@physik.tu-chemnitz.de}
\author{Hong-liu Yang}
 \thanks{corresponding authors}
% \email{hongliu.yang@physik.tu-chemnitz.de}
\affiliation{%
Institute of Physics,
Chemnitz University of Technology, D-09107 Chemnitz, Germany 
}

\date{\today}
\begin{abstract}
A frequently encountered situation in the study of delay systems is that
 the length of the delay time changes with time, which is of relevance in 
 many fields such as optics, mechanical machining, biology or physiology.
%  In this study we use a simple map system to investigate
%the influence of the fluctuating delay time on the system
%dynamics. 
A characteristic feature of such systems is that the dimension
of the system dynamics collapses due to the fluctuations of delay
times. In consequence, the support of the long-trajectory attractors of this kind of systems
is found being fractal in contrast to the fuzzy attractors in most
random systems.
\end{abstract}


%\keywords{Suggested keywords}

\maketitle

{\it Introduction.}
In ubiquitous natural and laboratory situations the action of time delayed signals is an essential ingredient to
understand the system dynamics. For instance, in optical and physiological systems \cite{optic,glass,neural}, 
a finite transmission speed usually leads to delayed reactions to signals from remote origins and/or 
time shifts among signals taking different paths. 
In mechanical engineering the rotation of the workpiece is the origin of time delay effects \cite{mach},  
and in feedback control methods time delayed signals are utilized 
to stabilize the system behavior \cite{feedback}. 

A frequently encountered situation which was rarely addressed previously \cite{tde}, is that the length of the delay time often varies
 with time. A biological example is that the reproduction cycle of animals fluctuates with the environment change.
 In mechanical engineering, vibrations of the machine tool and/or the workpiece may
  also change the length of the delay time \cite{mach2}. As a modern control scheme,
  the time variation of delays has been found to enhance the control efficiency \cite{fb2}. 
  
  The relevance of fluctuating delay times for all these phenomena
  motivates us to gain a better understanding of its influence on the system dynamics.
   In this contribution, we will
  start from a very simple map system where the delay time takes only the value of one
or zero discrete time steps. High dimensional systems with longer fluctuating delays show similar behavior \cite{wang,radons}.

A characteristic feature of such systems is that the system dynamics collapses to a low dimensional subspace due to the fluctuation of delay
times. One would expect that such a dimensional collapse will have some interesting consequences on the system dynamics. 
As an example, we show in this paper its influence on the fractal properties of the system attractors.
%Following the numerical exploration of these rich dynamics, we provide an analytical calculation of the Lyapunov exponent for
%the case of delayed Bernoulli shift map. The obtained analytical estimation of the Lyapunov exponent agrees very well with the
%numerical results. Moreover, the analytical calculation demonstrates the existence of a dense set of singularities in the
%Lyapunov exponent which corresponding to certain superstable nonperiodic attractors.

{\it Models.}
For simplicity, we focus on the map system
\begin{equation}
x_{t+1}=1-ax_{t}^2+(b+k\varepsilon _{t})x_{t-1}  \label{ex1}
\end{equation}
where $k$ and $b$ are constants and the time dependent random variable $\epsilon_{t}$ takes only value $0$ or $1$. 
The probability of $\epsilon_{t}$ taking the value 1 is denoted as $p$. For a parameter
setting with $b=0$ the length of the delay time of our system can vary randomly between one or zero time steps with 
the system dynamics simply switching between the Henon map ($\epsilon_t=1$) and the logistic
map ($\epsilon_t=0$). To illustrate the importance of the delay time variation and the resulting dimensional collapse 
a comparison is made with the well-studied case of the random Henon map in \cite{rbo} with
$b\neq 0$ and $\epsilon_{t}$ being an uniform noise in the interval $[0,1]$, where the delay time is constantly 2. 
In the following the parameter $a$ is fixed as $a=1.1$ if it is not stated otherwise.

{\it Numerical results.}
As depicted in \cite{rbo} to study the effect of randomness on attractors of a system like Eq.(\ref{ex1}) one may consider two different situations, either the {\it snapshot attractor} formed by an ensemble of
points at a single instant of time with all points starting from different initial conditions and evolving under a same given realization of the external noise,
 or the {\it long-trajectory attractor} formed by a long
trajectory segment starting from a single initial condition and evolving under a given realization of the external noise. The main issue addressed here will be
the difference between the long-trajectory attractors of the random delay case and the random Henon case.

\begin{figure}
\includegraphics*[scale=0.3]{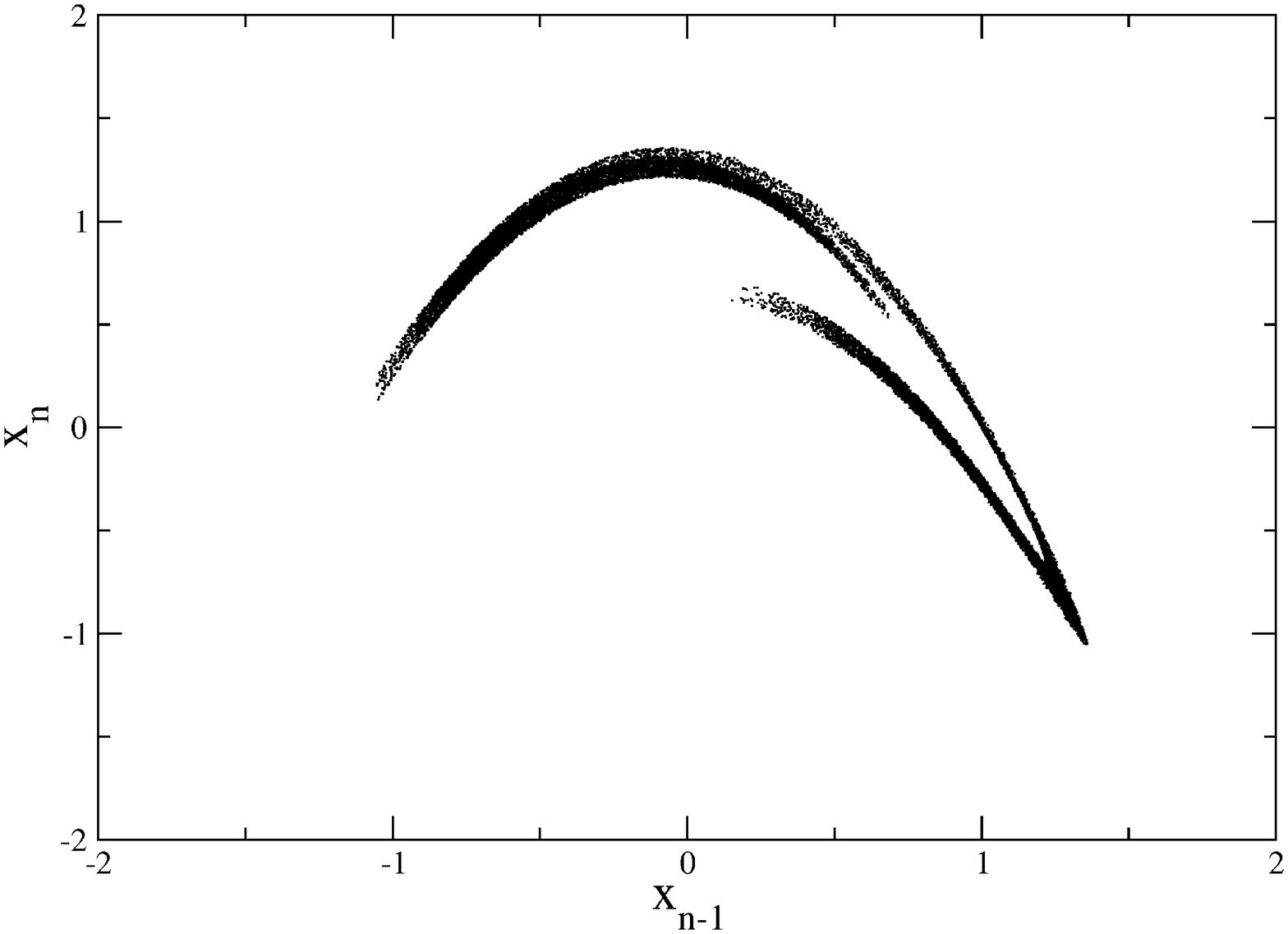}
\includegraphics*[scale=0.3]{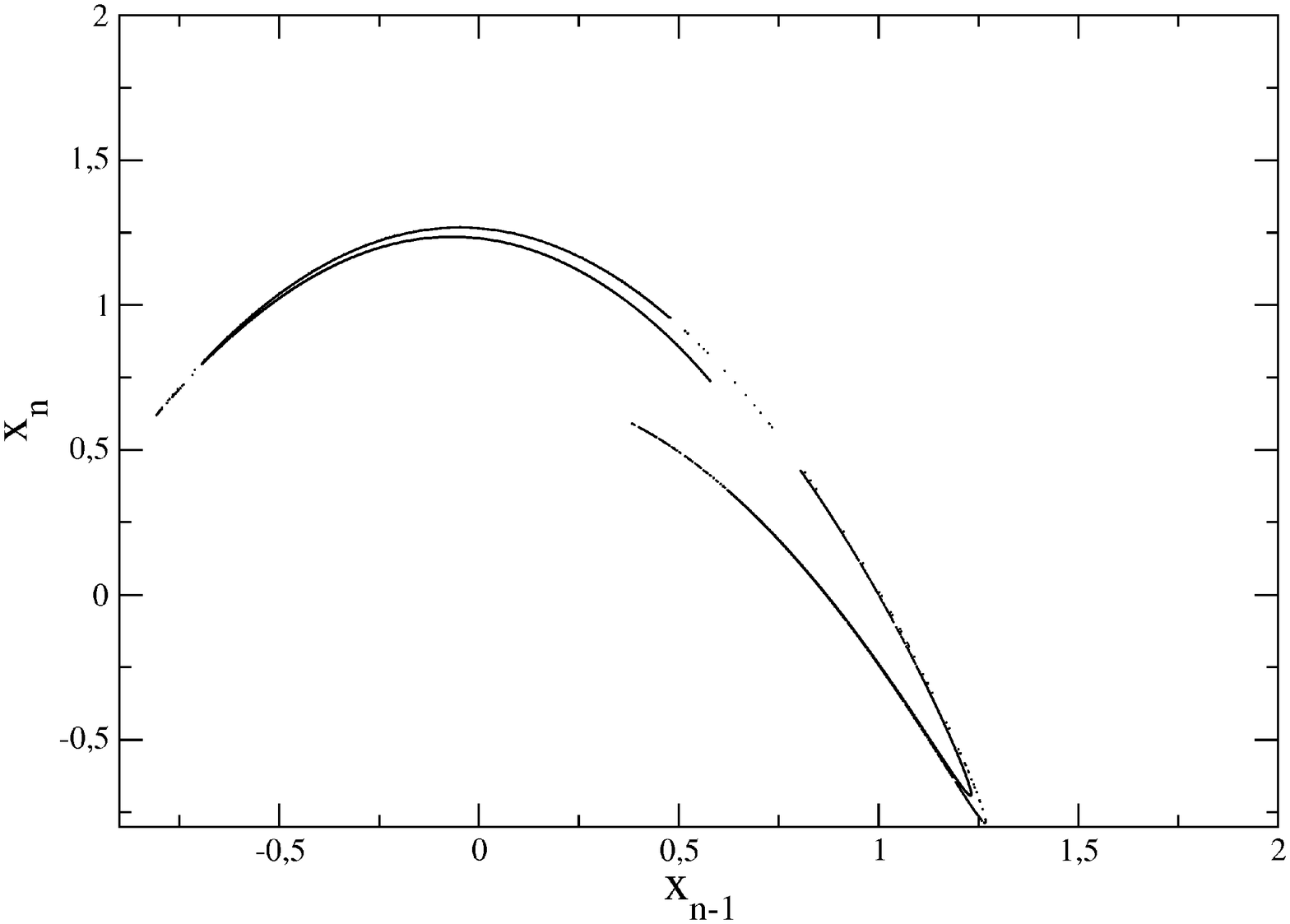}
\includegraphics*[scale=0.3]{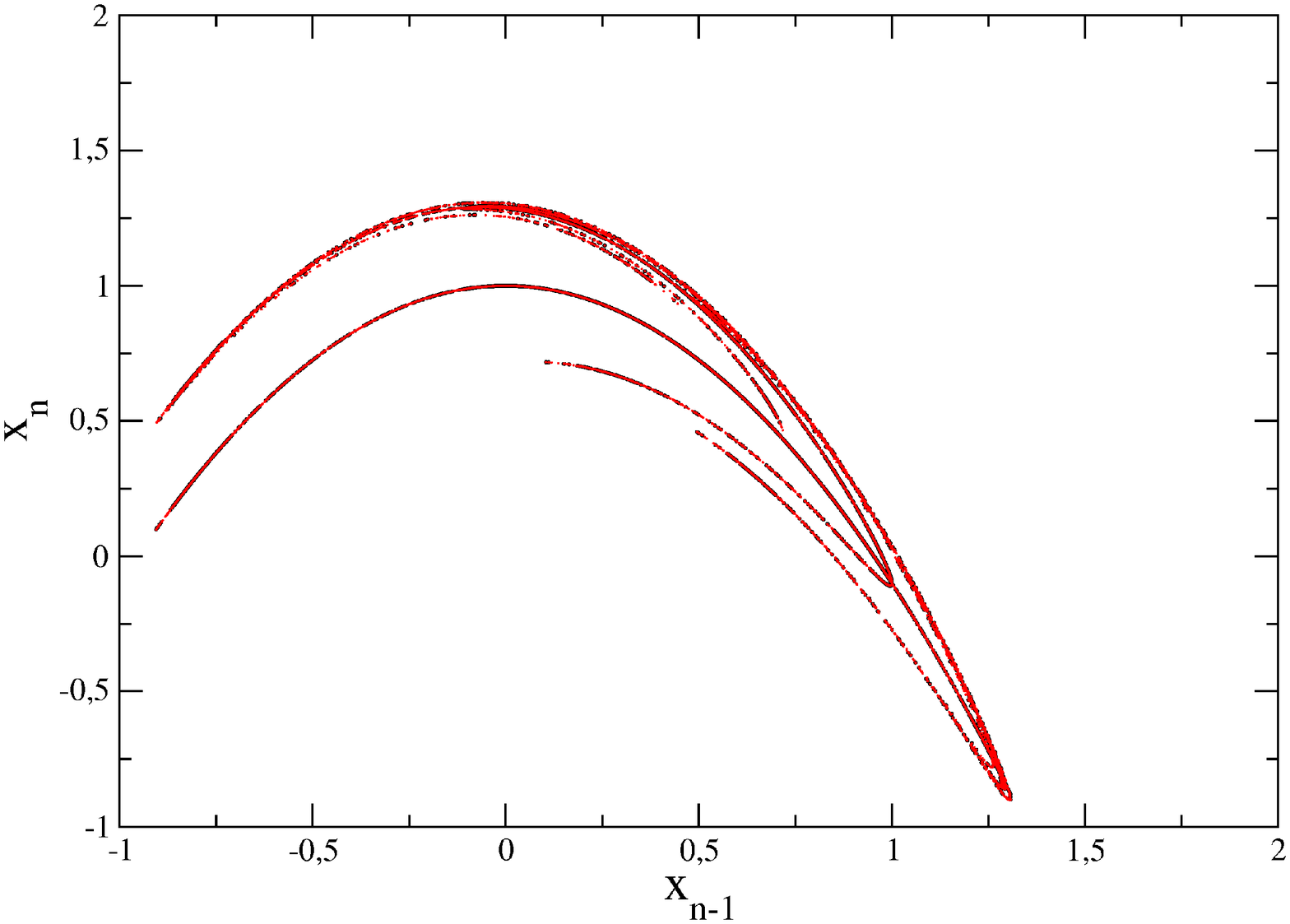}
\includegraphics*[scale=0.3]{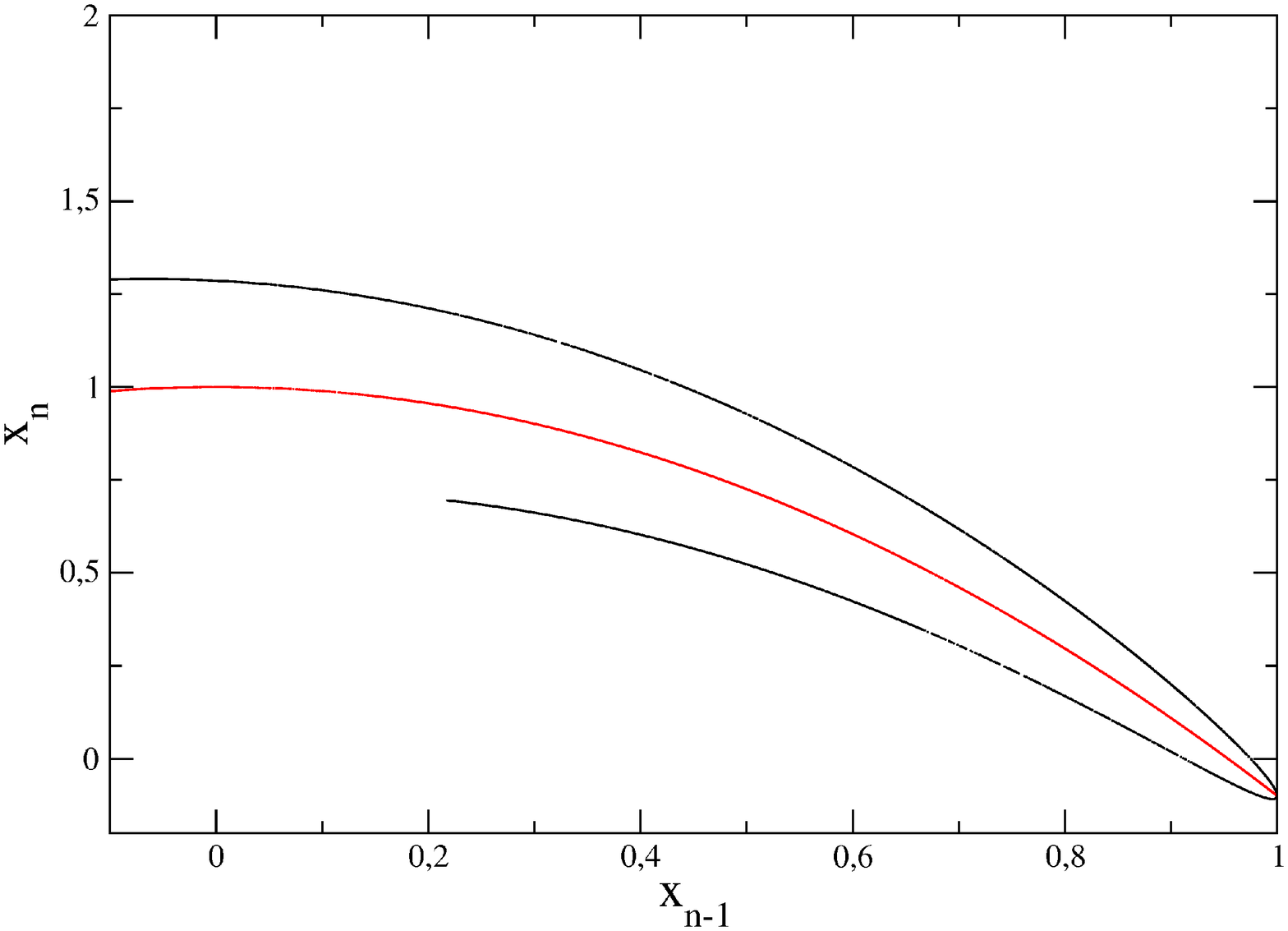}
\caption{\label{fig:fig1} 
Long-trajectory and snapshot attractors, respectively, for the random Henon map with $b=0.3$, $k=0.05$ and $p=0.5$ 
(a)-(b) and for the random delay system with $b=0$, $k=0.3$ and $p=0.5$ (c)-(d).
In (c) two runs with different initial conditions and different realizations of the noise are shown. In (d) two attractors at different instants of time are shown.
The corresponding capacity and information dimensions are listed in Tab. I.
}
\end{figure}

It was shown \cite{rbo} that for the random Henon case $b\neq 0$ the long-trajectory attractor is 
a fuzzy attractor with a {\it smooth} density of points while
the corresponding snapshot attractor may be fractal. An example with $b=0.3$, $k=0.05$ is presented in Fig.\ref{fig:fig1} (a) and (b).
In contrast, for the random delay case $b=0$ we find that the long-trajectory attractor turns out to be fractal in consequence of the delay time variation and the dimensional collapse. 
In Fig.\ref{fig:fig1} (c) and (d) the long-trajectory and snapshot attractors are shown for a random delay case with $b=0$, $k=0.3$ and $p=0.5$. 
Moreover, as demonstrated the long-trajectory attractors obtained are identical for two runs with
different initial conditions and different realizations of the noise $\epsilon_t$. This is different from previously investigated random systems 
where in general different realizations of external noise lead to different attracting sets lying in
different regions of the phase space. The snapshot attractor of the random delay case is time dependent and turns out 
to be at probability 1 a one-dimensional curve (see Fig.\ref{fig:fig1} (d)). 

\begin{figure}
\includegraphics*[scale=0.3]{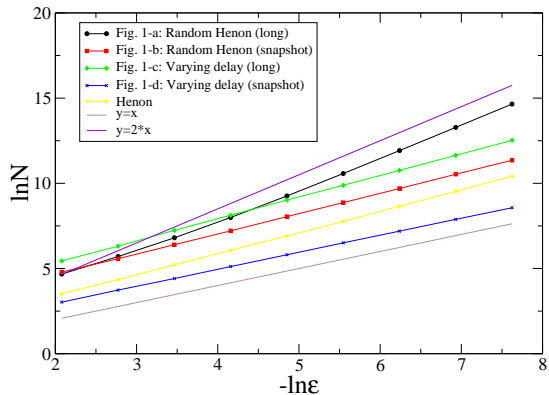}
\caption{\label{fig:fig2} 
Capacity dimension of the snapshot and long-trajectory attractors for the random delay map and the random Henon map as presented in \ref{fig:fig1}.
}
\end{figure}

To show quantitatively the difference in the fractal properties between the random delay case and the 
random Henon case we calculated the capacity and information dimension of the attractor via the box-counting algorithm.
The variation of the number of non-empty boxes $N(l)$ versus the box size $l$ is plotted in Fig.\ref{fig:fig2}. By definition the slope of the log-log plot of $N(l)$ versus $1/l$ gives the capacity dimension $D_0$ of the
attractor. Four curves in Fig.\ref{fig:fig2} correspond to the four attractors shown in Fig.\ref{fig:fig1}. To guide the eyes, three lines of slope $1$, $2$ and $1.26$ (the capacity dimension of the Henon attractor) 
are plotted in
the same figure. One can easily see from the plot that the capacity dimension of the long-trajectory attractor of the random Henon case and the random delay case is very close to $2$ and $1.26$ 
respectively. This 
confirms our above observation from Fig.\ref{fig:fig1} that the long trajectory attractor of the random Henon case is a fuzzy attractor with a smooth density of points while the long trajectory attractor of the 
random delay case is a fractal. The capacity dimension of the typical snapshot attractor of the random Henon case and the random delay case is $1.14$ and $1$ respectively, as expected from the observation that the former
attractor is fractal while the latter is a one-dimensional curve.

Moreover, we calculated also the information dimension $D_1$ of the four attractors shown in Fig.\ref{fig:fig1}. Results are given in Tab.1. Notice that the information dimension of the long trajectory attractor
for the random delay case is close to $1$.

\begin{figure}
\includegraphics*[scale=0.3]{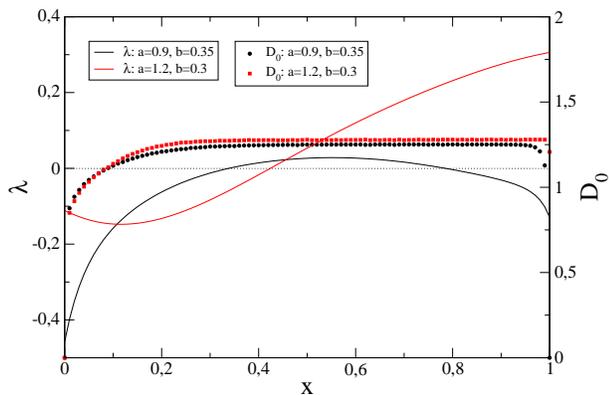}
\caption{\label{fig:fig3} 
(Color online) Lyapunov exponent $\lambda$ (curves) and the capacity dimension $D_0$ (symbols) of the long-trajectory attractor vs. probability $p$ 
for two random-delay cases with $b=0$. 
Notice the existence of transitions between stable and chaotic states.
}
\end{figure}

To characterize the dynamical behavior of the system we calculate the Lyapunov spectrum of the random delay case via the standard method \cite{benettin}. In general the system has one Lyapunov exponent of finite value and
the other of value $-\infty$ due to the dimension collapse. The Kaplan-Yorke dimension $D_{\lambda}$ therefore has the value $1$,
 consistent with the obtained value $D_1\simeq 1$ for the long trajectory attractor
as depicted in Tab.1. 
Furthermore, as shown in
Fig.\ref{fig:fig3} the leading Lyapunov exponent of the system can be positive or negative depending on the probability $p$.
 The capacity dimension $D_0$, however, stays constant in the regime $p>0.2$ irrespective of the value of $p$. 
 The smaller value of $D_0$ obtained in the regime
 $p<0.2$ is a numerical artifact due to an insufficient number of phase points used.
  Simulations show indeed that the obtained value of $D_0$ for $p<02$ increases with the
 number of phase points used and the regime of constant $D_0$ expands correspondingly.
An additional interesting point in Fig. \ref{fig:fig3} is that a randomly temporal switching between two nonchaotic dynamics with trivial attractors
results in a chaotic dynamics holding a fractal
attractor (e.g. for
$a=0.9$, $k=0.35$ and $p=0.5$).
Without considering the positive Lyapunov exponent corresponding to the external noise $\epsilon_t$ the fractal attractor of the random delay case with a negative Lyapunov exponent may be viewed as a strange
nonchaotic attractor \cite{sna}.

 \begin{table}[ht]
\centering
 \caption{Calculating the capacity and information dimension of the four attractors in Fig. 1}
\begin{tabular}{ccc}
 & $D_0$ & $D_1$\\
\hline
Fig.1-a & $1.95$ & $1.85$\\
Fig.1-b & $1.14$ & $1.06$\\
Fig.1-c & $1.27$ & $1.09$\\
Fig.1-d & $1.00$ & $0.84$\\
%Henon & $1.162$ & $1.123$\\
\end{tabular}

\end{table}

{\it Dimension collapse.}
A characteristic feature of the random delay system Eq.(\ref{ex1}) is that the length of the delay time fluctuates between 0 and 1 randomly. 
It is therefore convenient for our analysis to partition the evolution of the system
dynamics into segments having either pure delay time 0 or 1, wherein the system dynamics simply corresponds to the logistic or the Henon map,
 respectively. For
simplicity, we will denote them as 0- or 1- segment respectively.
 Whenever the system switches from a 1-segment to a 0-segment, the phase
point representing the dynamic evolution of the system will fall {\it immediately} on the parabola $z_{n+1}=1-az_n$ and 
the evolution of the system dynamics sticks in such a one-dimensional subspace during the 0-segment. 
This is what we called {\it dimension collapse}. 
At the end of the 0-segment, the system enters a
1-segment and the dynamics is now just the Henon map without randomness until the appearance of 
the next 0-segment.     
It is obvious that the target of the collapses is not influenced by the history of the system dynamics or the delay variations. 
The dynamic evolution of the random delay system Eq.(\ref{ex1}) can therefore be viewed as the
re-injections to the parabola during the 0-segments and the iterations of the parabola under the Henon dynamics during the 1-segments. 
The long trajectory attractor of the system can thus be roughly viewed as the union of the parabola and its
$n$-fold image under the Henon dynamics where $n$ is the possible length of the 1-segments and $n=1$, $2$, $\cdots$, $+\infty$.

An expression of the natural measure $\rho^*$ of the attractor can be worked out accordingly.
Based on the above discussions one may decompose $\rho^*$ in the following way
\begin{equation}
\rho^* =\sum_{n=0}^{\infty } \rho_{Ln} =\rho_{L0}+\sum_{n=1}^{\infty } \rho_{Ln} \label{m1}
\end{equation}
where $\rho_{L0}$ is the part of the natural measure living on the parabola given by the logistic map and 
$\rho_{Ln}$ represents the part of the nature measure whose support is the $n$-fold iteration of the parabola under
the Henon dynamics.
On the other hand a self-consistent equation for $\rho^*$ can be written down as
\begin{equation}
\rho^* =p\hat{H}\rho^* +(1-p)\hat{L}\rho^* \label{m2}
\end{equation}
 which is a Frobenius-Perron like equation for our random system Eq.(\ref{ex1}). Here $\hat{H}$ and $\hat{L}$ denote the evolution operators for the density of
  the logistic and the Henon map respectively.
Inserting (\ref{m1}) in (\ref{m2}) reads
\begin{equation}
\rho_{L0}+\sum_{n=1}^{\infty } \rho_{Ln}  =(1-p)\hat{L}\rho^*+ p\sum_{n=0}^{\infty }\hat{H}\rho_{Ln} \label{m3}.
\end{equation}
Identifying terms on two sides of (\ref{m3}) which have the same phase space support leads to
\begin{eqnarray}
\rho_{L0}&=&(1-p)\hat{L}\rho^*; \label{m4}  \\
\rho_{Ln}  &=&p^{n}\hat{H}^{n}\rho_{L0} \thinspace\thinspace\thinspace\thinspace\thinspace\thinspace\text\thinspace\thinspace\thinspace{for}\thinspace\thinspace\thinspace n>0  \label{m5}.
\end{eqnarray}
Inserting (\ref{m5}) in (\ref{m1}) generates a new expression of $\rho^*$
\begin{equation} 
\label{m7} 
\rho^* =\sum_{n=0}^{\infty } p^n\hat{H}^{n}\rho_{L0}.
\end{equation}
As a consistence check one can insert (\ref{m4}) in (\ref{m7}), it reads
\begin{eqnarray}
\rho^* &=&(1-p)\sum_{n=0}^{\infty } p^n\hat{H}^{n} \hat{L}\rho^* \\
  &=&(1-p) (\mathbf{1}-p\hat{H})^{-1} \hat{L}\rho^*  \label{m8},
\end{eqnarray}
which is just a simple reformulation of the Frobenius-Perron equation given in (\ref{m2}).

Eq.(\ref{m4}) indicates that $\rho_{L0}$ is the projection of the natural measure $\rho^*$ on the parabola of the logistic map. It therefore 
has a continuous density on this one dimensional subject for
the used parameter. 
As discussed the support of $\rho_{Ln}$ is the $n$-fold iteration of the parabola under the action of the Henon map and it approaches the Henon attractor
 as $n$ goes to $+\infty$. Dimensional collapse means that the dimension of $\rho_{Ln}$ is smaller than the complete dimension of the phase space which in consequence implies
 that the iterations of $\rho_{Ln}$ under the action of the Henon map are overlapped only at measure zero points. One would thus expect
from Eq.(\ref{m5}) that the capacity dimension $D_0$ of the (long trajectory) attractor is smaller than the dimension of the phase space while the information
dimension $D_1$ should be determined by $\rho_{L0}$ and has the value $1$. 
This explains the appearance of a fractal attractor in our random delay
system Eq. (\ref{ex1}) and the above obtained values of $D_1$ of the corresponding attractors. It is necessary to point out that the capacity dimension 
$D_0$ of the (long trajectory) attractor is in general different from that of the corresponding Henon attractor
 (see the case $a=0.9$, $b=0.35$ in Fig. \ref{fig:fig3}), although they are very close if the corresponding Henon dynamics
is chaotic. 

Moreover as can be seen from Eq.(\ref{m5}), while other Renyi dimensions may vary with the probability $p$
the capacity and information dimensions of the long-trajectory attractor are independent of $p$ (see Fig. \ref{fig:fig3} for numerical confirmation).

Note that the random occurrence of segments of logistic map dynamics in our random delay system interrupts the previous Henon map evolution and starts a new
Henon evolution with a different point on the parabola of the logistic map as the initial condition. 
The randomness of the delay variation can only determine when the reinjection happens and prepare a new initial condition for the next Henon evolution.
 This is obviously irrelevant for the global
geometrical properties of the long-trajectory attractor. Therefore different realizations of external noise $\epsilon_t$ would lead
to the same long trajectory attractor as also indicated by Eq.(\ref{m5}) (see Fig.1 (c)). 
The dynamical evolution of phase point surely depends on the realization of the
external noise, see for example the variation of the shape of the snapshot attractor. In this sense the snapshot and long-trajectory attractors contain
different information of the system dynamics. This is assumed due to the nonstationary of the evolution rule of random systems.

%One more point to be noticed is that the collapse of the system dynamics to the parabola is immediately, i.e. there is no relaxation process. The importance of this feature of dimensional collapse to
%the appearance of the fractal attractor can be shown clearly comparing to a seemly similar situation of the random Henon case. For the random Henon case $b=0$, one may tune the parameters $a$ and $k$ to get a
%situation that the system has a chaotic or fixed-point attractor for $\epsilon_t=0$ and $1$ respectively. Although the (asymptotic) attractor for $\epsilon_t=0$ is of low dimension than that for
%$\epsilon_t=1$ as in the random delay system the long-trajectory attractor is still a fuzzy attractor with a continuous density of points (not shown). The essential difference is that the relaxation time to the
%low dimensional attractor is nonzero for the random Henon case.   

As can be seen from above analysis the collapse of the phase space dimension works as a new mechanism of generating fractal attractor in random delay systems like (\ref{ex1}). 
The collapse ensures that all segments of Henon evolutions starts synchronously from the same low dimensional subspace, the parabola of the logistic map.
 Such a synchronization prevents the
randomness to blur the fine structure of the attractors generated from the Henon evolutions.
%In addition, it is essential that the dimension of the collapsed subspace (the parabola) is lower than that of the asymptotic attractor of the constant long delay dynamics (the
%Henon attractor). Such a analysis can be generalized to the high dimensional case straightforwardly.
%Consider now a
%high dimensional map system with the delay time switching randomly between $n$ and $m$ with $n<m$. Assume that the map with constant delay time $m$ has a 
%attractor of the capacity dimension
%$D_m$ and the projection of the attractor to the $n$-dimensional space has the dimensional $D_n^{(p)}$. 
%Due to a similar analysis one would expect that the capacity dimension of the long trajectory attractor of this random delay
%system is simply $\max\{D_n^{(p)},D_m\}$. 
%It is also clear from our above analysis that other
%properties of the given dynamics such as the local or global contraction is irrelevant for the geometrical properties of the attractor \cite{ifs}.

{\it Summary.}
As a first step towards a deep understanding of the behavior of systems with fluctuating delay time, we studied a simple map system with
 either one or two step memory of the past
states. Numerical simulations and supplementary arguments showed that the dimension collapse of the random delay system has 
some very interesting consequences in the system
dynamics for instance the fractal properties of the attractor.

\begin{acknowledgments}
Support from the Deutsche Forschungsgemeinschaft (DFG Grant No. Ra416/6-1) is gratefully acknowledged. 

\end{acknowledgments}

\end{document}